\documentclass{moriond}

\usepackage{graphicx}
\usepackage{amsmath,amssymb}
\usepackage{xspace}

\usepackage{multicol,caption}

\newcommand{\pp}{\mbox{$pp$}\xspace}

\newcommand{\akt}{\mbox{anti-$k_{t}$}\xspace}

\newcommand{\RpPb}{\mbox{$R_{p \mathrm{Pb}}$}\xspace}

\newcommand{\Tpb}{\mbox{$T_{\mathrm{Pb}}$}\xspace}

\newcommand{\avgTpb}{\mbox{$\langle \Tpb \rangle$}\xspace}
\newcommand{\sqn}{\mbox{$\sqrt{s_{\mathrm{NN}}}$}\xspace}
\newcommand{\sqs}{\mbox{$\sqrt{s}$}\xspace}

\newcommand{\pPb}{\mbox{$p$+Pb}\xspace}
\newcommand{\pT}{\mbox{$p_\mathrm{T}$}\xspace}
\newcommand{\TeV}{\mbox{Te\kern -0.1em V}\xspace}
\newcommand{\GeV}{\mbox{Ge\kern -0.1em V}\xspace}

\newcommand{\Photo}{}

\newenvironment{Figure}
  {\par\medskip\noindent\minipage{\linewidth}}
  {\endminipage\par\medskip}

\begin{document}

\title{Measurement of the nuclear modification factor for high-$\pT$ charged hadrons in \pPb collisions with the ATLAS detector}
\author{Petr Balek (for the ATLAS Collaboration)}

\address{Department of Particle Physics and Astrophysics, Faculty of Physics, Weizmann Institute of Science,  234 Herzl Street, Rehovot 7610001, Israel}

\maketitle

\begin{abstract}
The charged hadron spectra in \pPb and \pp collisions at $\sqn=\sqs=5.02$\,\TeV are measured with the ATLAS experiment at the LHC. The measurements are performed with \pPb data recorded in 2013 with an integrated luminosity of 25\,nb${}^{-1}$ and \pp\ data recorded in 2015 with an integrated luminosity of 28\,pb${}^{-1}$. The \pPb results are compared to \pp spectra, presented as a ratio of transverse momentum distributions in the two systems scaled by the number of binary nucleon--nucleon collisions, the nuclear modification factor \RpPb. The study of \RpPb allows a detailed comparison of the collision systems in different centrality intervals and in a wide range of transverse momentum. It is shown that the nuclear modification factor does not have any significant deviation from unity in the high transverse momentum region.

\hspace*{1cm}

\noindent \textit{keywords}: proton--lead collision, nuclear modification factor
\end{abstract}

\begin{multicols}{2}

\section{Introduction}

The deep inelastic scattering experiments revealed~\cite{Arneodo:1992wf} that the parton distribution functions in nuclei differ from an incoherent superposition of nucleons. This includes phenomena of shadowing, anti-shadowing~\cite{Salgado:2011wc} and the EMC effect~\cite{Aubert1983275}. Proton--ion collisions at the LHC allow probing nuclei structure at scales that were not experimentally accessible and searching for previously unobserved phenomena. 

The modification of charged particle production in \pPb collisions is quantified by the nuclear modification factor ($\RpPb$) defined as

\begin{equation}
\RpPb = \frac{1}{\avgTpb}\frac{1/N_\mathrm{evt}~\mathrm{d}^2 N_{{\footnotesize p}\mathrm{\footnotesize Pb}} / \mathrm{d}y \mathrm{d}\pT}{\mathrm{d}^2\mathrm{\sigma}_\mathrm{\footnotesize pp} / \mathrm{d}y \mathrm{d}\pT},
\end{equation}
\noindent where $\avgTpb$ is the nuclear thickness function accounting for increased flux of partons per collision in \pPb collisions and it is estimated using Glauber model~\cite{Glauber}, $N_\mathrm{evt}$ is the number of \pPb events, $\mathrm{d}^2 N_{{\footnotesize p}\mathrm{\footnotesize Pb}} / \mathrm{d}y \mathrm{d}\pT$ is the differential yield of charged particles in \pPb collisions measured differentially in transverse momentum of tracks \pT\ and rapidity $y$~\footnote{ATLAS uses a right-handed coordinate system with its origin at the nominal interaction point (IP) in the center of the detector and the $z$-axis along the beam pipe. The $x$-axis points from the IP to the center of the LHC ring, and the $y$ axis points upward. Cylindrical coordinates $(r,\phi)$ are used in the transverse plane, $\phi$ being the azimuthal angle around the beam pipe. For the \pPb\ collisions, the incident Pb beam travelled in the $+z$ direction. The pseudorapidity is defined in laboratory coordinates in terms of the polar angle $\theta$ as $\eta=-\ln\tan(\theta/2)$. Rapidity is defined $y=0.5\ln{\left[(E+p_{z})/(E-p_{z})\right]}$, where $E$ is the energy and $p_{z}$ is the longitudinal momentum.},
and $\mathrm{d}^2\mathrm{\sigma}_\mathrm{\footnotesize pp} / \mathrm{d}y \mathrm{d}\pT$ is the differential charged particle production cross section in \pp collisions. 
In the case a \pPb collision is a superposition of several \pp collisions, $\RpPb$ should not differ from unity.

Using high statistics of 2013 \pPb data, the CMS collaboration measured the inclusive charged particle production~\cite{cms_rppb} that showed an enhancement of approximately 30\% for particles with $\pT\gtrsim20$\,\GeV with respect to a baseline, derived from the \pp data taken at lower and at higher center-of-mass energies. No \pp data at $\sqrt{s}=5.02$\,\TeV\ were available at that time. Preliminary results from ATLAS~\cite{ATLAS-CONF-2014-029}, which also rely on a similar interpolation, showed an enhancement compatible with the CMS results. The ALICE experiment measured \RpPb in a lower $\pT$ region~\cite{alice_rppb}. Trends shown in ALICE results do not support, but also do not fully exclude the enhancement seen by CMS. In the recent CMS measurement~\cite{Khachatryan:2016odn} using real \pp data at $\sqrt{s}=5.02$\,\TeV from 2015 data taking period, the enhancement is reduced compared to the original result~\cite{cms_rppb}.

\section{Analysis}

This proceeding reports the results that use \pp data collected in 2015 with the ATLAS detector~\cite{Aad:2008zzm} at $\sqs=5.02$\,\TeV with a total integrated luminosity of 25\,pb${}^{-1}$. Details can be found in Ref.~\cite{ATLAS-CONF-2016-108}. Measured charged particle cross section is compared to the \pPb spectra presented in Ref.~\cite{ATLAS-CONF-2014-029} in order to provide a direct comparison of charged particle productions in both systems using the data measured at the same center-of-mass energy.

This measurement is performed using the ATLAS tracking system (inner detector), calorimeter system, trigger system and data acquisition system. A particle in the inner detector typically crosses four Pixel layers, 4 double sided micro-strip layers (SCT) and around 36 straw tubes (TRT). The whole inner detector covers $|\eta|<2.5$ and is immersed in a 2\,T axial magnetic field. The calorimeter system consists of an electromagnetic calorimeter covering $|\eta|<3.2$, a hadronic calorimeter covering also $|\eta|<3.2$ and two forward calorimeters with a coverage of $3.2<|\eta|<4.9$. The calorimeter system provides 10 interaction lengths of material.

Events were recorded with a minimum bias (MB) trigger and with several jet triggers. The MB trigger required a track to be reconstructed in the inner detector. The jet triggers used the \akt algorithm~\cite{anti-kt} with radius parameter of $R=0.4$ and with $p_\mathrm{T}^\mathrm{jet}$ thresholds of 30, 40, 50, 60, 75 and 85\,\GeV.

To study the effects of the detector response, Monte Carlo (MC) simulations were produced using the PYTHIA 8 event generator~\cite{pythia8}. The detector response was then simulated by GEANT4~\cite{geant,simul_infra} and reconstructed and analyzed in the same way as the data. The MC samples were produced in different exclusive kinematic intervals of leading jet $p_\mathrm{T}^\mathrm{jet}$, allowing sufficient statistics over wide range of track $\pT$.

Charged-particle tracks are reconstructed in the ATLAS inner detector over pseudorapidity region $|\eta|<2.5$ and over full azimuth angle. At least one hit is required in one of the two inner layers of the Pixel detector. If the track passed an active module in the innermost layer a hit in this layer is required. Furthermore, a track has to have no more than two holes in the Pixel and SCT detectors together. A hole is defined by the absence of a hit predicted by the track trajectory. In addition, for a track with $|\eta|<1.65$, the presence of at least 9 Pixel and SCT hits together is required. For a track with higher $|\eta|$, the requirement is raised to 11 hits. In case there are less than 10 Pixel and SCT hits, no holes are allowed along the track. To ensure a good matching to the vertex, a $\pT$-dependent requirement on $d_0$ is imposed on tracks, where $d_0$ represents the distance of the closest approach to the vertex in the transverse direction.

Tracks in events recorded by any jet trigger are further required to be matched to a jet within $\Delta R = \sqrt{\Delta^2 \eta + \Delta^2 \phi} < 0.4$ and to have $\pT$ $\leq$$1.3\times p_\mathrm{T}^\mathrm{jet}$. This condition accounts for energy conservation and possible jet energy mismeasurement. 

No jet is considered, unless its energy is high enough to allow the trigger to be fully efficient. However, if a jet trigger with higher threshold could also be fully efficient, the jet from lower threshold triggered event is not considered neither. Thus, to have a fully efficient trigger is a necessity condition but not sufficient condition. This approach prevents double counting of the same events if they were recorded by more than one trigger.

Several corrections are applied in order to remove detector effects and reconstruction biases. First, the measured charge particle spectra are corrected for secondary and fake tracks. Those are either tracks matching to secondary particles or they are spurious combinations of hits not associated with a single particle. This correction, as well as all other corrections, is estimated as a function of rapidity ($y$) and $\pT$. Rapidity is calculated assuming all particles have mass of pion and there is a correction dealing with this bias. The correction for secondary and fake tracks is estimated to be around 1\% at low $\pT$ and as much as 7\% at high $\pT$ and high $|y|$. The second correction is performed for the $\pT$ resolution. Iterative Bayesian unfolding~\cite{BayesUnf} is used to correct for possible $\pT$ mismeasurement. Stable results are achieved after 2 iterations. The unfolding introduces a change of about 5\% at high $\pT$, while the effect for low $\pT$ tracks is much smaller. The next correction is for the track reconstruction efficiency. It is estimated after excluding fake and secondary tracks from consideration. The efficiency is around 85\% at mid-rapidity and low $\pT$, increasing to 90\% at high $\pT$. Efficiencies at higher $|y|$ are reduced by no more than 15\%. Finally, a last correction is applied to account for the fact that the pion mass is used for all tracks to calculate the rapidity in all previous corrections as well as in measured distributions. This gives a bias for other particles (mostly kaons and protons). The correction is determined using MC simulations with the true mass of each particle. It is up to 1\% at $\pT\approx5$\,\GeV\ and decreases with increasing $\pT$.

Systematic uncertainties are evaluated by varying their individual sources and comparing the results to that of the default analysis. Main contribution comes from the unfolding procedure. It is around 1\% at low $\pT$, however at high $\pT$ it reaches up to 7\%. The uncertainty covers possible distortions due to limited statistics in the migration matrices used by the Bayesian unfolding. Another significant uncertainty arises from the uncertainty of the matching between reconstructed tracks and generated particles. It is also 7\% at the highest $\pT$ and high $|y|$. Variation of the track selection requirements introduces an uncertainty no more than 5\%. Uncertainty on the luminosity is 5\%. Other sources of the uncertainties, such as detector material, correction for fake and secondary tracks or track reconstruction efficiency do not exceed 4\%.

\section{Results}

The corrected charged particle cross section measured in \pp collisions at $\sqs=5.02$\,\TeV is shown in Fig.~\ref{fig:spectra}. It is measured in the same rapidity interval as \pPb\ spectra~\cite{ATLAS-CONF-2014-029} which are shown in the same figure as well. The centrality class of a particular \pPb\ event is based on the total transverse energy measured in the Pb-going side of the forward calorimeter.

Figure~\ref{fig:ratio} shows comparisons of several distributions to the interpolated \pp reference. This reference was used previously for the estimation of $\RpPb$ together with the \pPb spectra~\cite{ATLAS-CONF-2014-029}. The data, represented by black crosses, agree well with the interpolation at low $\pT$ but  they are about 30\% higher at high $\pT$. Comparisons to alternative interpolation methods and to the PYTHIA cross section are also shown. They agree within a few per cent at low $\pT$ and within 20\% at high $\pT$.

\begin{Figure}
	\centering
	\includegraphics[width=\linewidth]{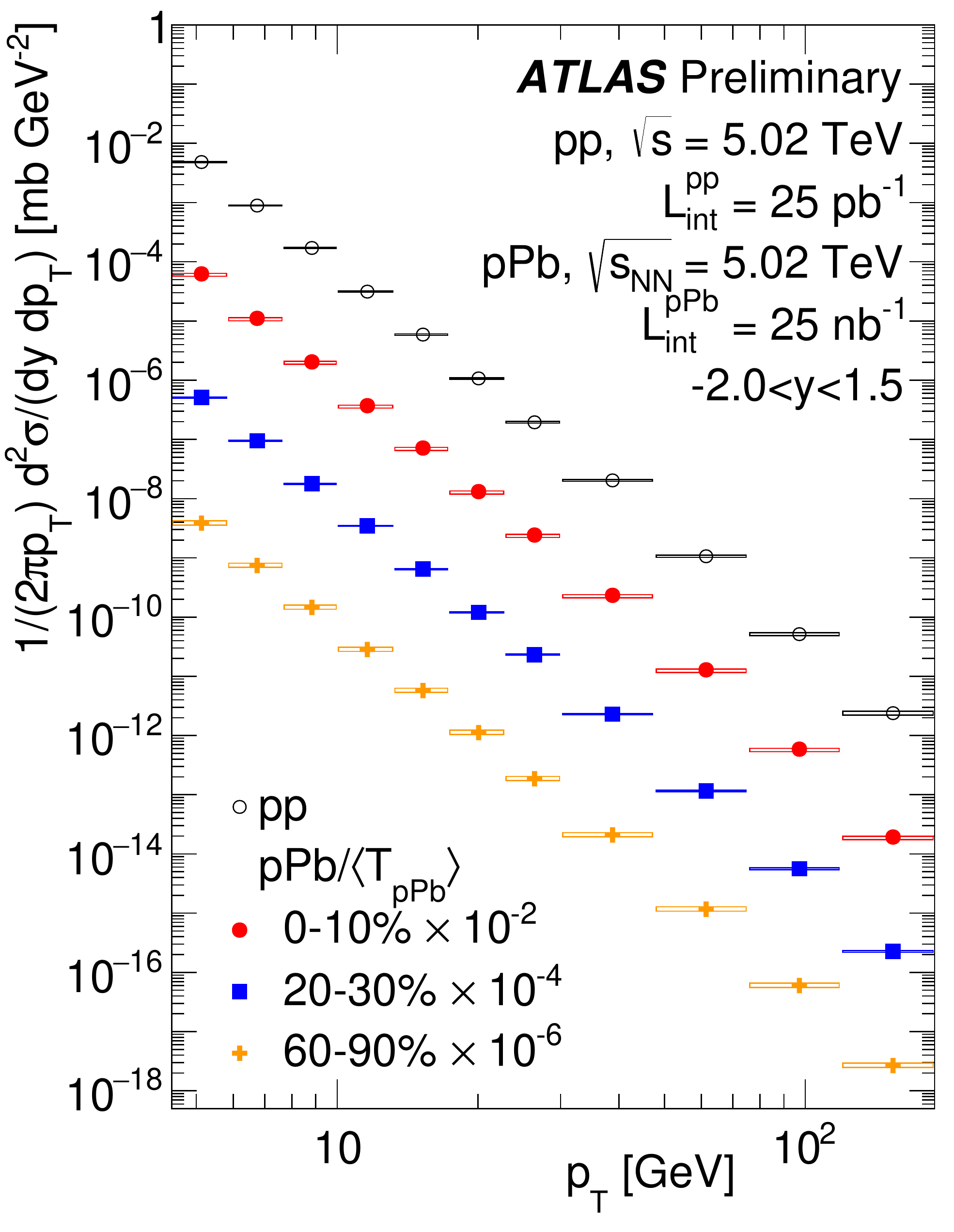}
	\captionof{figure}{Charged particle production cross section measured in \pp collisions at $\sqs=5.02$\,\TeV~\protect\cite{ATLAS-CONF-2016-108}. The \pPb were taken from Ref.~\protect\cite{ATLAS-CONF-2014-029}. Boxes represent systematic uncertainties; the statistical uncertainties are smaller than marker size.}
	\label{fig:spectra}
\end{Figure}

\vspace{0.3cm}

\begin{Figure}
	\centering
	\includegraphics[width=\linewidth]{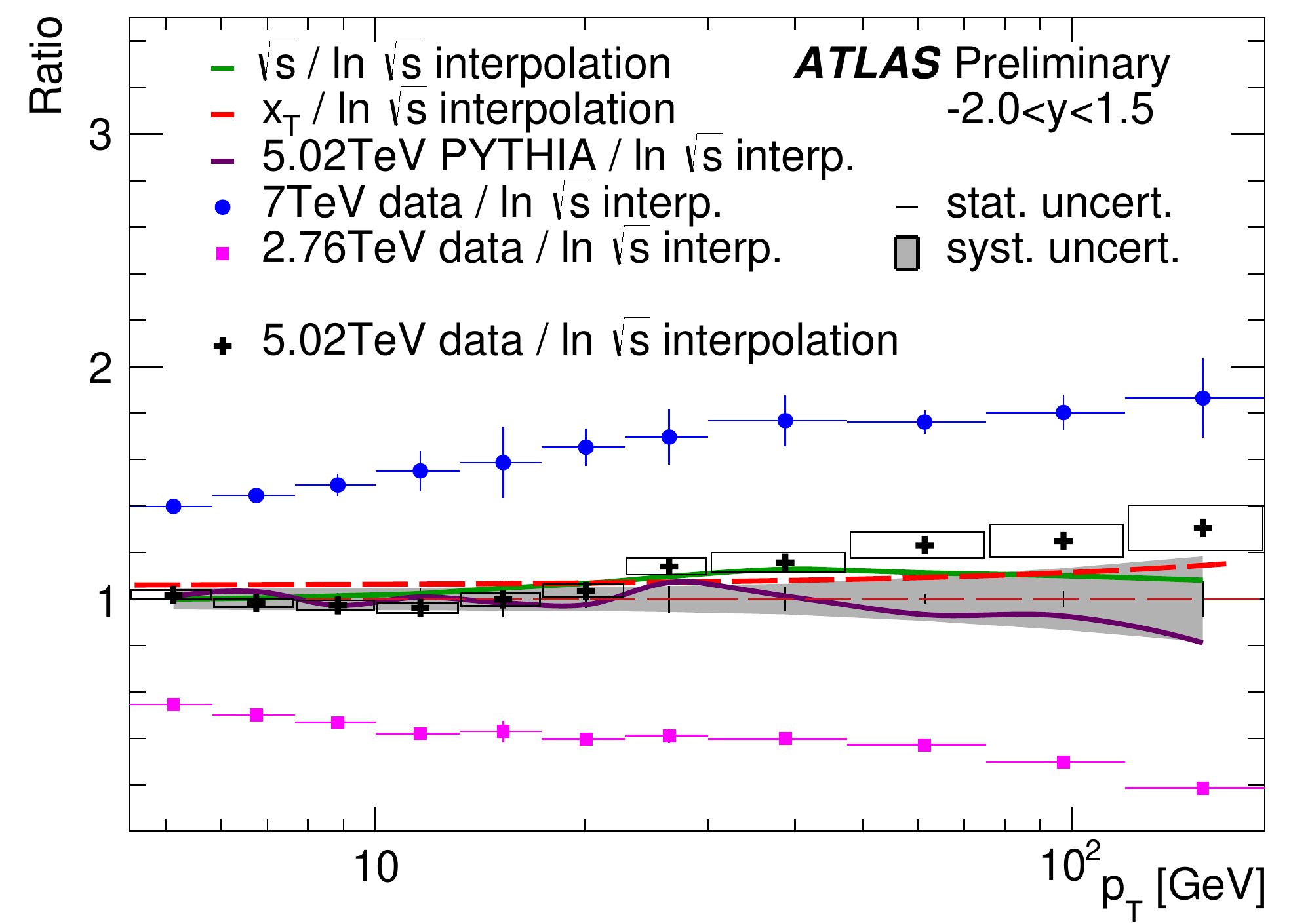}
	\captionof{figure}{Ratios to the interpolated \pp reference~\protect\cite{ATLAS-CONF-2016-108}. The ratios of the data measured at 2.76, 5.02 and 7\,\TeV are represented by circles, crosses and squares, respectively. The boxes around crosses depict the systematic uncertainty of the presented 5.02\,\TeV measurement. Green line shows ratio of $\sqrt{s}$ interpolation, red line of $x_\mathrm{T}$ interpolation, and magenta line of the PYTHIA cross section. The gray band and vertical lines around unity represent systematic and statistic uncertainty of the interpolated \pp reference, respectively.}
	\label{fig:ratio}
\end{Figure}

Figure~\ref{fig:rppbMB} shows the nuclear modification factor $\RpPb$ as a function of $\pT$ in the 0--90\% centrality interval. The $\pT$ dependence is substantially reduced compared to the previous ATLAS estimate relying only upon the interpolation, that is also shown. The $\RpPb$ does not show any significant deviation from unity in the whole $\pT$ region.

Figure~\ref{fig:rppb} shows $\RpPb$ in three centrality intervals. The most central interval 0--10\% shows an increase toward lower $\pT$, while the peripheral interval shows a decrease. This is consistent with previous ATLAS measurement~\cite{rppb_zhenia_paper}, that is also shown. There is no centrality dependence of the $\RpPb$ at high $\pT$.

\vspace{0.2cm}

\begin{Figure}
	\centering
	\includegraphics[width=\linewidth]{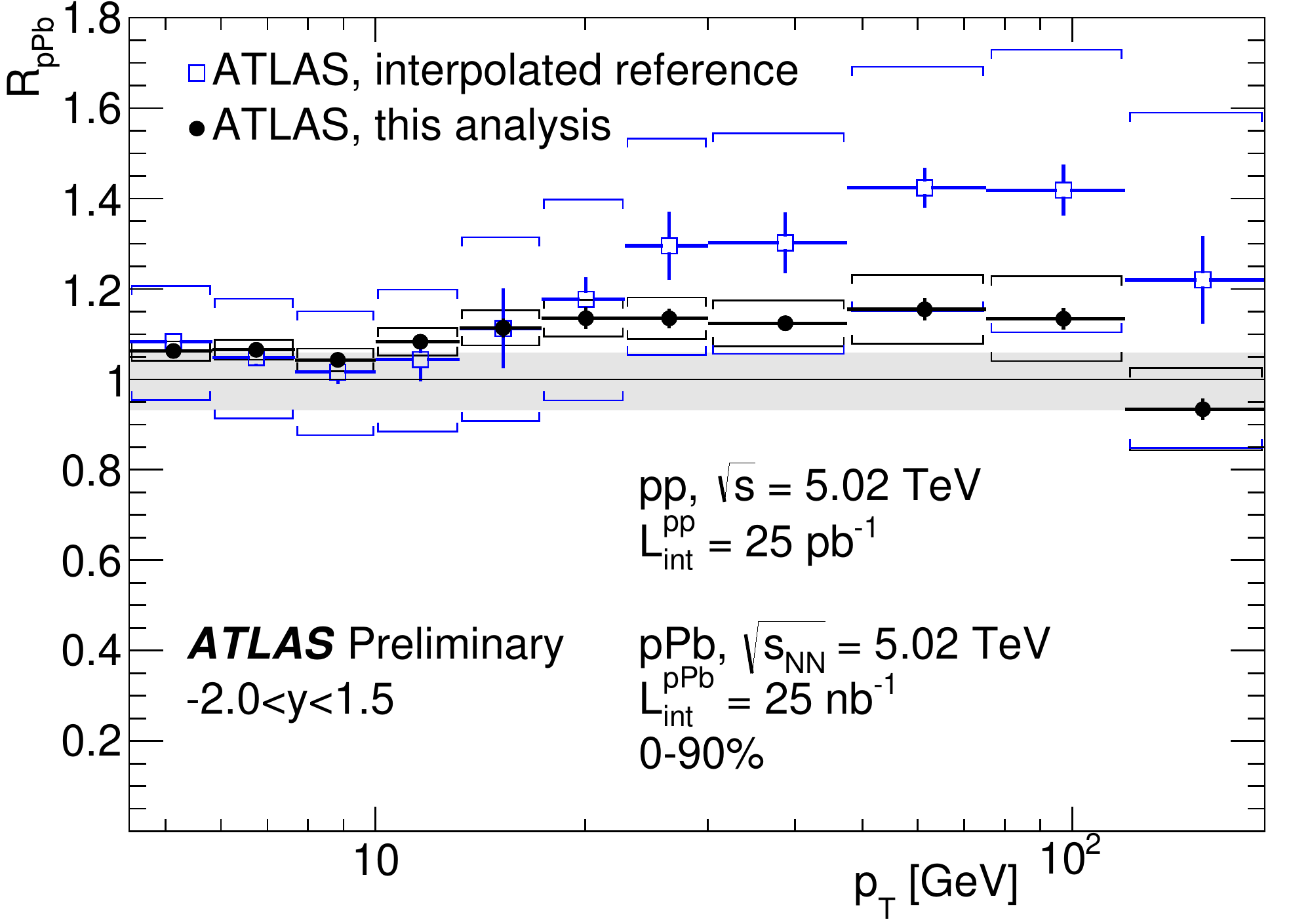}
	\captionof{figure}{Nuclear modification factor $\RpPb$ measured in the 0--90\% centrality interval~\protect\cite{ATLAS-CONF-2016-108}. Statistical uncertainties are shown with vertical lines and brackets represent systematic uncertainties, with an exception of the fully correlated systematic uncertainty that is shown with the gray band around unity. The result of the previous analysis~\protect\cite{ATLAS-CONF-2014-029} relying upon interpolation is shown with blue squares.}
	\label{fig:rppbMB}
\end{Figure}

\vspace{0.4cm}

\begin{Figure}
	\centering
	\includegraphics[width=\linewidth]{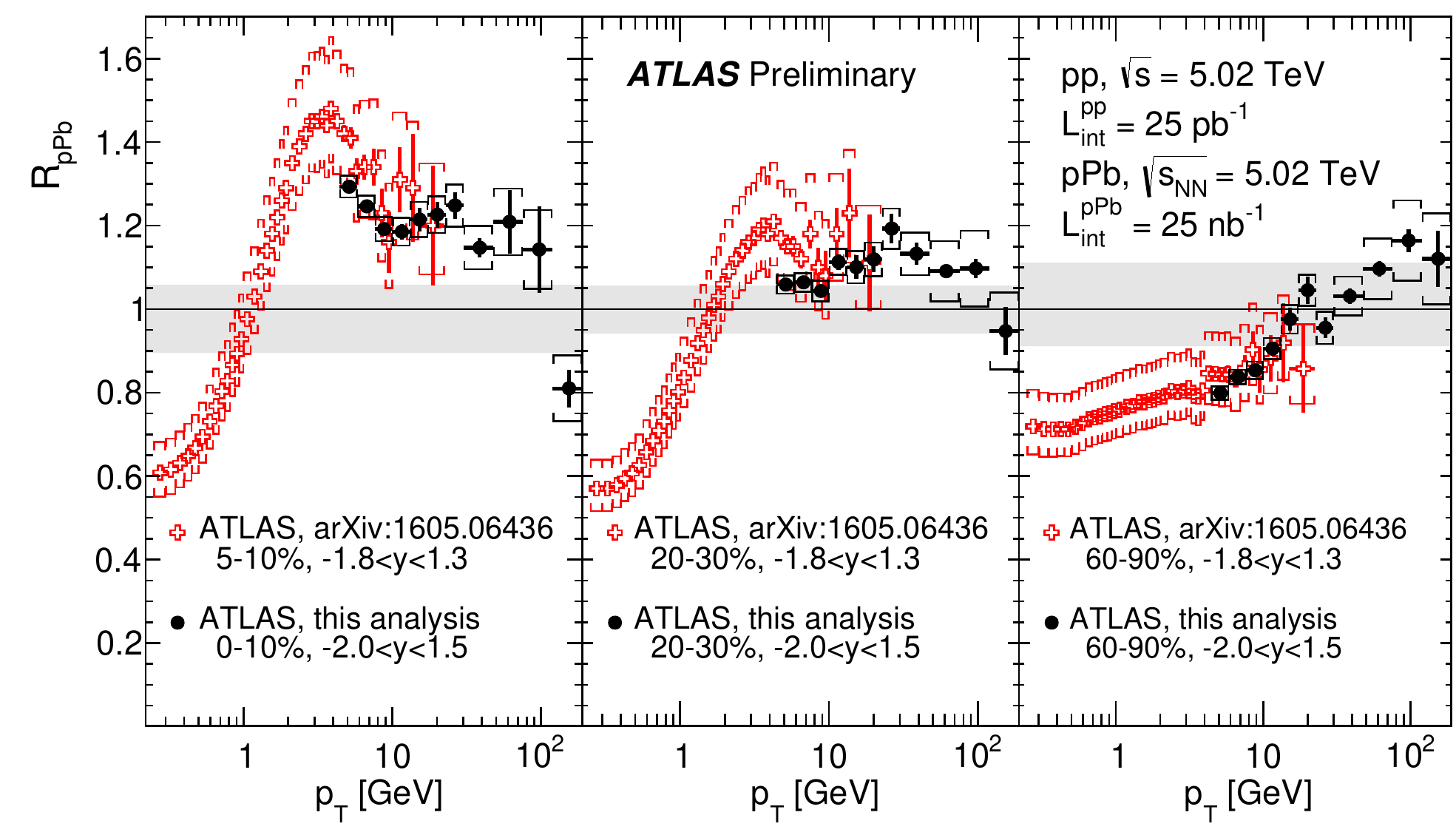}
	\captionof{figure}{Nuclear modification factor $\RpPb$ measured in the three centrality intervals: 0--10\%, 20--30\% and 60--90\%~\protect\cite{ATLAS-CONF-2016-108}. Statistical uncertainties are shown with vertical lines and brackets represent systematic uncertainties, with an exception of the fully correlated systematic uncertainty that is shown with the gray band around unity. The results of the previous analysis~\protect\cite{rppb_zhenia_paper} focusing on low $\pT$ region are shown with red crosses.}
	\label{fig:rppb}
\end{Figure}

\vspace{0.3cm}

\section{Summary}

Charged particle production in \pp collisions at $\sqs=5.02$\,Tev has been presented. It is used as a reference for the measurement of the nuclear modification factor $\RpPb$.
The $\RpPb$ shows a strong centrality dependent behaviour at low $\pT$. At high $\pT$ there is no centrality dependence and $\RpPb$ is consistent with unity. This is in agreement with recent CMS result~\cite{Khachatryan:2016odn}.

\section*{Acknowledgements}
This research is supported by the Israel Science Foundation (grant 1065/15) and by the MINERVA Stiftung with the funds from the BMBF of the Federal Republic of Germany.

\bibliographystyle{elsarticle-num}
\bibliography{Balek_P}
\end{multicols}
\end{document}